# Discovery of a metallic oxide with ultralow thermal conductivity

Jianhong Dai[1,2,*], Zhehong Liu[1,2,*], Jialin Ji[3], Xuejuan Dong[1], Jihai Yu[4], Xubin Ye[1,2], Weipeng Wang[1], RiCheng Yu[1,2], Zhiwei Hu[5], Huaizhou Zhao[1,2,§], Xiangang Wan[4], Wenqing Zhang[6,7,‡], & Youwen Long[1,2,8,†]

[1]Beijing National Laboratory for Condensed Matter Physics, Institute of Physics, Chinese Academy of Sciences, Beijing 100190, China

[2]School of Physics, University of Chinese Academy of Sciences, Beijing 100049, China

[3]Materials Genome Institute, Shanghai University, Shanghai 200444, China.

[4]Department of Physics, Nanjing University, Nanjing 210093, China

[5]Max Planck Institute for Chemical Physics of Solids, NöthnitzerStraße 40, 01187 Dresden, Germany

[6]Department of Materials Science and Engineering and Shenzhen Institute for Quantum Science & Technology, Southern University of Science and Technology, Shenzhen, Guangdong 518055, China.

[7]Shenzhen Municipal Key Lab for Advanced Quantum Materials and Devices and Guangdong Provincial Key Lab for Computational Science and Materials Design, Southern University of Science and Technology, Shenzhen, Guangdong 518055, China.

[8]Songshan Lake Materials Laboratory, Dongguan 523808, Guangdong, China

# Abstract

A compound with metallic electrical conductivity usually has a considerable total thermal conductivity because both electrons and photons contribute to thermal transport. Here, we show an exceptional example of iridium oxide, $Bi_3Ir_3O_{11}$, that concurrently displays metallic electrical conductivity and ultralow thermal conductivity approaching 0.61 W m$^{-1}$ K$^{-1}$ at 300 K. The compound crystallizes into a cubic structural framework with space group $Pn$-3. The edge- and corner-sharing $IrO_6$ octahedra with a mixed $Ir^{4.33+}$ charge state favor metallic electrical transport. $Bi_3Ir_3O_{11}$ exhibits an extremely low lattice thermal conductivity close to the minimum limit in theory owing to its tunnel-like structure with filled heavy atoms Bi rattling inside. Theoretical calculations reveal the underlying mechanisms for the extraordinary compatibility between metallic electrical conductivity and ultralow thermal conductivity. This study may establish a new avenue for designing and developing unprecedented heat-insulation metals.

Thermal conductivity is a fundamental physical property of solid-state materials. Motivated by the widespread applications [1-3], many materials with distinct thermal transport properties, such as the super-high thermal conductivity of diamond (1000-2200 W m$^{-1}$ K$^{-1}$) [4] and the extremely low thermal conductivity of silica aerogel (0.02 W m$^{-1}$ K$^{-1}$) [5], have been developed. In principle, for a compound with metallic electrical transport, a portion of the total thermal conductivity ($\kappa_{tot}$) can be attributed to the electronic contribution based on the Wiedemann-Franz (WF) law [6]. Here, metallic electrical transport is defined based on a positive temperature ($T$) derivative of resistivity ($\rho$): $d\rho/dT > 0$. Consequently, metallic materials tend to exhibit excellent heat dissipation owing to the high density of heat-carrying electrons. To date, only a few metallic oxides, such as $Na_{0.77}MnO_2$ [7], $Cd_2Os_2O_7$ [8], $NaCo_2O_4$ [9, 10], and $Bi_2Sr_2Co_2O_y$ [11], show relatively low $\kappa_{tot}$ close to 1.5-3.0 W m$^{-1}$ K$^{-1}$ at 300 K. Most of these metallic oxides crystalline into a 2-dimentioanl structure. In comparison, little is known for a 3-dimensional (3D) isotropic material that shows metallic electrical conductivity and low thermal conductivity of less than 1.0 W m$^{-1}$ K$^{-1}$ at room temperature.

In the field of thermoelectric technology, a series of semiconducting materials with the features of "electron crystal and phonon glass" have been developed in the last decade [1-3]. Some complex compounds with tunnel- or cage-like crystal structures were prepared to obtain phonon glass-like materials with reduced lattice thermal conductivity ($\kappa_{lat}$). In the primitive unit cells of such systems, there are large numbers of atoms with atomic rattling motions in the tunnels or cages, as observed in skutterudites [12,13] and clathrates [14]. For example, in the Bi-doped 3D semiconducting skutterudite $CoSb_3$, the $\kappa_{tot}$ measured at 450 K can be suppressed to 1.7 W m$^{-1}$ K$^{-1}$ due to the Bi atomic rattling-induced anharmonic phonon glass behavior [15]. Based on an anharmonic lattice, if some transition-metal ions with mixed valence states and weak electronic correlations (e.g., $5d$ electrons) can be introduced, it is possible to concurrently realize metallic electrical conductivity and ultralow thermal conductivity. Following this scenario, a new $5d$ transition-metal oxide $Bi_3Ir_3O_{11}$ with a 3D crystal structure and mixed $Ir^{4.33+}$ valence state was designed and prepared under high-pressure and high-temperature conditions. The compound possesses a cubic $Pn$-3 symmetry with 68 atoms in a primitive unit cell, forming a tunnel-like crystal structure with rattling heavy Bi atoms. As a result, metallic conductivity and extremely low thermal conductivity (~0.61 W $m^{-1}$ $K^{-1}$ at 300 K) occur in this new material, making it promising for many potential applications, for example in thermoelectric devices [16], battery electrodes [17], supercapacitors [18], electro-catalysis [19], and shielding coatings [20].

Detailed experimental methods and theoretical calculations used in this study are described in the Supplementary Material [21]. Fig. 1A shows the X-ray diffraction (XRD) pattern of $Bi_3Ir_3O_{11}$ measured at room temperature. The results of the Rietveld analysis [22] indicate that the compound crystallizes into a cubic crystal structure with space group $Pn$-3. In this 3D structural framework, there are two special Wyckoff positions for Bi: $4b$ (0, 0, 0) for $Bi_1$ and $8e$ ($x, x, x$) for $Bi_2$; one special position 12 $g$ ($x$, 0.75, 0.25) for Ir; and three positions for O, i.e., $12f$ ($x$, 0.25, 0.25) for $O_1$, $8e$ ($x, x, x$) for $O_2$, and $24h$ ($x, y, z$) for $O_3$. The refined structural parameters, such as the lattice constant, atomic positions, bond lengths, and angles, are summarized in Supplementary Tables S1 and S2. During structural refinement, the occupancy factors for Bi and Ir were both close to 100% (see Table S1), ruling out the atomic antisite disorder in $Bi_3Ir_3O_{11}$. As shown in Fig. 1B, the schematic structure of $Bi_3Ir_3O_{11}$ indicates that each iridium is surrounded by six ligand oxygens ($2O_1 + 4O_3$), forming an $IrO_6$ octahedron. Two adjacent octahedra build up an edge-sharing dioctahedron, which are perpendicularly connected to each other by shared corners. Therefore, the edge- and corner-sharing $IrO_6$ octahedra provide pathways for the carrier electrical transport of Ir ions. Similar to $Bi_3Ru_3O_{11}$ [23], in $Bi_3Ir_3O_{11}$, each $Bi_1$ atom is surrounded by eight O atoms (see Supplementary Fig. S1A) with an average $Bi_1$-O bond length of 2.427 Å. However, there are nine coordinated O atoms for each $Bi_2$ with an average bond length of 2.603 Å (see Fig. S1B). Along the [111] direction (see Fig. 1C), tunnel-like $Ir_2O_{10}$ frameworks give rise to rather large interspaces. The $Bi_2$ atoms are loosely bound at approximately the centers of these interspaces, and therefore are thermally unstable. In a distorted $IrO_6$ octahedron, the Ir-O distance varies from 1.85(1) to 2.09(6) Å, with an average value of 1.99 Å. Compared with the Ir-O bond length for $Ir^{4+}$ (2.03 Å) and $Ir^{5+}$ (1.96 Å) ions in an $IrO_6$ octahedron, as shown in $La_2MgIrO_6$ [24] and $Cd_2Ir_2O_7$ [25], respectively, a mixed valence state of $Ir^{4.33+}$ is assigned to the current $Bi_3Ir_3O_{11}$ ($Ir^{4+}$ by 33% and $Ir^{5+}$ by 67%). This is in good agreement with the X-ray absorption spectroscopy measurements of Ir as shown in Fig. S2 [26,27], suggesting the stoichiometric chemical composition of $Bi_3Ir_3O_{11}$. A high-angle annular dark-field (HAADF) image was acquired along the [110] direction of $Bi_3Ir_3O_{11}$ to further demonstrate the refined structure; the results are shown in Fig. 1D. The bright dots, representing the Bi and Ir atom columns, corresponding to the constructed crystal structure as shown in its inset image.

Fig. 2A shows the temperature dependence of the resistivity of $Bi_3Ir_3O_{11}$. The resistivity value observed at 300 K was $11.4 \times 10^{-5}$ Ω m. Moreover, as the temperature decreased to 2 K, the resistivity decreased smoothly with $d\rho/dT > 0$, confirming the metallic electrical transport property; this was expected from the mixed $Ir^{4.33+}$ charge state mentioned above. Some other iridates, such as $Cd_2Ir_2O_7$ [25] with an $Ir^{5+}$ state and $Ba_2Ir_3O_9$ [28] with a mixed $Ir^{4.67+}$ state, also showed metallic conductivity with a resistivity scale similar to that observed in $Bi_3Ir_3O_{11}$. At lower temperatures, the electrical transport of the former two follows the Fermi liquid behavior ($\rho$ proportion to $T^2$). In contrast, the resistivity of $Bi_3Ir_3O_{11}$ deviates from this behavior and exhibits linear temperature dependence below 30 K (see the inset of Fig. 2A). This may imply different electronic correlated effects in the iridium oxides.

In accordance with the metallic electrical behavior, a small Seebeck coefficient was observed for $Bi_3Ir_3O_{11}$. As presented in Fig. 2B, the Seebeck coefficient gradually increased with increasing temperature, and a maximum value of 7.1 μV K$^{-1}$ occurred up to 60 K. Then, the Seebeck coefficient slightly decreased upon heating and became nearly constant above 200 K. Fig. 2C shows the specific heat $C_p$ as a function of temperature for $Bi_3Ir_3O_{11}$. Like the resistivity data, the specific heat also monotonically varies with the temperature from 2 to 300 K. This means that there is no structural or long-range magnetic order taking place over the temperature range, which is consistent with the magnetic susceptibility and magnetization measurements, as shown in Fig. S3, where only paramagnetic behavior was observed at 2-300 K. Considering the metallic and nonmagnetic nature, both electrons and phonons contributed to the $C_p$ of $Bi_3Ir_3O_{11}$. At lower temperatures (< 20 K), the $C_p$ data could be well fitted (see the inset of Fig. 2C) using the function $C_p = \gamma T + f_D \cdot 9 N_A k_B \left(\frac{T}{\theta_D}\right)^3 \int_0^{\theta_D/T} \frac{x^4 e^x}{(e^x-1)^2} dx + f_E \cdot 3 N_A k_B \left(\frac{\theta_E}{T}\right)^2 \frac{e^{\theta_E/T}}{\left(1-e^{\theta_E/T}\right)^2}$ yields $\gamma$ = 11.4 mJ mol$^{-1}$ K$^{-2}$, as expected for a metallic system. Here, $N_A$ is the Avogadro number; $\theta_D$ and $\theta_E$ are the Debye and Einstein temperatures, respectively; and $f_D$ and $f_E$ are the atomic contribution ratios of vibrational modes per formula unit in the Debye and Einstein models, respectively. According to the fitting result, the Debye temperature was determined to be $\theta_D = 65.4$ K. For most transition-metal oxides, the values of $\theta_D$ are typically 200-500 K [29,30]. The unusually low Debye temperature of $Bi_3Ir_3O_{11}$ may be suggestive of anharmonic atomic vibrations, which can considerably reduce $\kappa_{lat}$.

Fig. 2D illustrates the temperature dependence of the thermal conductivity of $Bi_3Ir_3O_{11}$. In sharp contrast to the considerable value of $\kappa_{tot}$, as expected for a metallic compound, the current $Bi_3Ir_3O_{11}$ exhibits ultralow thermal conductivity. Specifically, from 3 to 300 K, the $\kappa_{tot}$ gradually increases from 0.0097 to 0.61 W m$^{-1}$ K$^{-1}$, which is comparable with the value of the liquid water (0.60 W m$^{-1}$ K$^{-1}$) at ambient conditions. Near the temperature corresponding to the thermopower kink at approximately 60 K, there was a curvature change in the thermal conductivity. Below this temperature, $\kappa_{tot}$ increased nearly linearly with the temperature. Above this temperature, however, $\kappa_{tot}$— in particular, $\kappa_{lat}$— increased only slightly upon heating.

According to the WF law, the electronic contribution to thermal conductivity can be expressed as $\kappa_{ele} = L_{eff} T/\rho$, where $L_{eff}$ denotes the effective Lorenz constant. For most metals and alloys, the universal value is $L_{eff} = L_0 = 2.45 \times 10^{-8}$ W $\Omega$ K$^{-2}$. Based on the experimentally measured $\rho$-$T$ data shown in Fig. 2A, we can obtain the temperature dependence of $\kappa_{ele}$ for $Bi_3Ir_3O_{11}$ (see Fig. 2D) by temporarily assigning $L_{eff} = L_0$. After subtracting $\kappa_{ele}$ from $\kappa_{tot}$, $\kappa_{lat}$ was obtained as a function of the temperature, as shown in Fig. 2D. In this case, the $\kappa_{ele}$ and $\kappa_{lat}$ obtained at 300 K for $Bi_3Ir_3O_{11}$ are 0.11 and 0.50 W m$^{-1}$ K$^{-1}$, respectively. On the other hand, according to the atomic number $n$ per unit volume, the Boltzmann constant $k_B$ and the phonon spectrum, Clarke [31] theoretically proposed the minimum thermal conductivity ($\kappa_{min}$) contributed from the phonon with the function of $\kappa_{min} = 0.76 k_B n^{\frac{2}{3}}(2\nu_t + \nu_l)$, where $\nu_t$ and $\nu_l$ are the transverse and longitudinal sound velocities, respectively. We obtained $\nu_t = 1.56 \times 10^3$ and $\nu_l = 2.82 \times 10^3$ m s$^{-1}$ by acoustic velocity measurement for Bi$_3$Ir$_3$O$_{11}$ at 300 K. In our material system, the value of $n$ is 0.0816 Å$^{-2}$, and the $\kappa_{min}$ for acoustic phonons was calculated to be 0.43 W m$^{-1}$ K$^{-1}$. The value of $\kappa_{lat}$ (0.50 W m$^{-1}$ K$^{-1}$) derived from experiment for Bi$_3$Ir$_3$O$_{11}$ is close to this calculated $\kappa_{min}$. Here, the contribution of the optical phonons to $\kappa_{lat}$ is negligible because of the low group velocity. The results also mean that the $\kappa_{ele}$ of Bi$_3$Ir$_3$O$_{11}$ still follows the WF law with $L_{eff}$ equal (or close) to $L_0$ despite the existence of some moderate electronic correlated effects, as will be shown later. This is essentially different from the strongly correlated oxide VO$_2$, where the value of $L_{eff}$ (= $0.11L_0$) is claimed to be far away from $L_0$ because of the absence of quasiparticles in a strongly correlated electron fluid where heat and charge diffuse independently [32].

We now discuss the origins of the metallic electrical conductivity and ultralow thermal conductivity of Bi$_3$Ir$_3$O$_{11}$. As is well known, introducing disorder, impurities, point defects or interfaces, and alloying have been proposed to suppress the lattice thermal conductivity. However, none of these effects was dominant in the current Bi$_3$Ir$_3$O$_{11}$. The peculiar crystal construction and electronic state of Ir ions with mixed valence states are responsible for the metallic electrical transport and ultralow thermal conductivity. The edge- and corner-sharing IrO$_6$ octahedral portion provides a rigid sublattice, which is attributed to the metallic electrical conductivity dominated by mixed $Ir^{4.33+}$ electrons. In addition, there are numerous tunnels surrounded by $Ir_2O_{10}$ units, and the heavy Bi atoms are located inside these tunnels. Bi atoms vibrate with a large amplitude inside the tunnels because they are weakly bound. These vibrations are incoherent (and are therefore called "rattling" [33]) and act as traps for acoustic phonons. This results in intensive phonon scattering and the phonon contribution to the thermal conductivity is significantly reduced, giving rise to the ultralow thermal conductivity of $Bi_3Ir_3O_{11}$.

To further understand the electronic properties of $Bi_3Ir_3O_{11}$, we performed first-principles calculations based on density functional theory. As shown in Fig. 3A, our calculations show a small band gap (approximately 26.0 meV) when the electron-electron correlated energy $U$ is not considered. As the importance of electronic correlations for $5d$ orbitals has been recently emphasized [34,35], we also carried out a generalized gradient approximation (GGA) + $U$ scheme[36] in consideration of the correlated effect and the varied parameter $U$ from 0 to 4.0 eV. Usually, increasing $U$ enlarges the bandgap. However, including $U$ reduces the band gap and the compound becomes metal (gap closed) when $U \geq 2.0$ eV, as shown in Fig. 3B. This can be attributed to the delicate interplay of strong spin-orbital coupling and moderate electronic correlations for the $5d$ electrons in $Bi_3Ir_3O_{11}$. We also calculated the Sommerfeld coefficient $\gamma$ based on the density of states at the Fermi level from the $U = 2.0 - 4.0$ eV (see Fig. S4). As an example, the numerical result at 3.0 eV is $\gamma = 6.39$ mJ mol$^{-1}$ K$^2$, which is comparable to the experimentally fitted result for the low-$T$ specific heat data mentioned above.

The molecular dynamic (MD) simulations were performed to get a deeper insight into the ultralow thermal conductivity of $Bi_3Ir_3O_{11}$. Fig. 3C shows the simulated trajectory of the atoms in the [100] direction at 300 K. The large displacements of $Bi_1$ and $Bi_2$ (particularly $Bi_2$ with remarkable anisotropy) in real space present incoherent disordered vibrations of atomic substructures, although all atoms retain their equilibrium sites as a whole. Due to the extensiveness of local vibration trajectory, the dynamic process of the $Bi_2$ atoms inside the structure are easy to visualize as the state of "rattling". Compared with $Bi_1$, $Bi_2$ shows a more intense expansion of the vibration trajectory, as the number of rattling $Bi_2$ atoms is four times larger than that of $Bi_1$ in the tunnel-like crystal structure. The intrinsically low $\kappa_{lat}$ is thus expected because the introduction of local vibration into the crystal structure produces local weak chemical bonds and leads to phonon resonance scattering. Smaller interatomic bonding and larger atomic masses result in an anharmonic oscillator because of the tunnel-like structure of $Bi_3Ir_3O_{11}$ with rattling heavy Bi atoms. This further introduces low-frequency optical phonons that significantly reduce the $\kappa_{lat}$.

We calculated the phonon spectra to study the contribution of the phonon perspective to low-temperature thermal conductivity. As shown on the left side of Fig. 3D, the acoustic waves converged on $\Gamma$, denoted as longitudinal and two transversal phonon branches. The most distinct feature of $Bi_3Ir_3O_{11}$, in terms of its lattice dynamics, is the low-frequency optic phonons. These overlap the frequency section of acoustic phonon, as low as 0.8 THz at the $\Gamma$ point. This frequency is much lower than that of most low-frequency optical phonons in other low-$\kappa_{lat}$ compounds, such as filled skutterudite

BaCo$_4$Sb$_{12}$ (13 THz) and CoSb$_3$ (15 THz) [37]. Therefore, the acoustic phonon frequency can be reduced by a large margin. The traditional interaction between phonons in low- $\kappa_{lat}$ materials is mainly represented by the nonlinear scattering of acoustic phonons. In some systems, such as that of skutterudite, the lattice thermal conductivity cannot be explained by traditional phonon interactions alone, and an additional resonant scattering process related to the random perturbation effect of packed atoms is required. In multiatomic unit cells, optical waves have a very low group velocity at the $\Gamma$ point in reciprocal space; this is suboptimal for transporting heat; therefore, the contribution to $\kappa_{lat}$ is rather small. However, low-frequency optical waves indirectly affect the thermal conductivity, mainly by coupling with acoustic phonons. For example, in some Cu-based uncaged structures with local vibrations, some Cu atoms have significantly larger displacement parameters, generating many low-frequency optical branches and thus giving rise to resonance scattering [38]. Similarly, the introduction of rattling Bi$_2$ atoms in Bi$_3$Ir$_3$O$_{11}$ causes low-frequency optical phonons, as shown on the right side of Fig. 3D. These low-frequency optical phonons interact with acoustic phonons and suppress their transmission. Consequently, the $\kappa_{lat}$ of the material was significantly reduced.

Figure 4A shows the temperature dependence of thermal conductivity for low-$\kappa_{tot}$ oxides including the metallic Na$_{0.77}$MnO$_2$ [7], Cd$_2$Os$_2$O$_7$ [8], NaCo$_2$O$_4$ [9,10], and Bi$_2$Sr$_2$Co$_2$O$_y$ [11] as well as the semiconducting/insulating BiCuSeO [39], LaCoO$_3$ [40], Ca$_3$Co$_4$O$_9$ [11], and LiCoO$_2$ [7]. In comparison, the current Bi$_3$Ir$_3$O$_{11}$ possesses the lowest $\kappa_{tot}$ in the whole temperature window we measured (3-300 K). Due to the limitation of low-$\kappa_{tot}$ metallic oxides, other more low-$\kappa_{tot}$ oxide materials with semiconducting/insulating behavior were also used to compare the thermal conductivity determined at 300 K. As shown in Fig. 4B, the value of the metallic $Bi_3Ir_3O_{11}$ locates at an intrinsically ultralow level among all oxides reported so far. This is very fascinating for a metallic oxide which possesses an ultralow thermal conductivity even compared with those of semiconductors and insulators. Moreover, $Bi_3Ir_3O_{11}$ crystallizes into a 3D cubic lattice; this rules out the anisotropic effects in structural symmetry, favoring potential applications for its ultralow thermal conductivity.

In summary, a novel ternary oxide, $Bi_3Ir_3O_{11}$, was obtained by high-pressure and high-temperature methods. This compound has a cubic *Pn*-3 crystal structure with tunnel-like frameworks in which heavy Bi atoms are loosely bonded. $Bi_3Ir_3O_{11}$ exhibits metallic electrical transport and a small Seebeck coefficient. First-principles calculations indicate that electronic correlated effects play an important role in the formation of metallic electronic band structures. Meanwhile, $Bi_3Ir_3O_{11}$ shows an ultralow thermal conductivity with a $\kappa_{tot}$ of less than 0.61 W m$^{-1}$ K$^{-1}$ over the entire temperature range of 3-300 K, which is the lowest in oxide materials with metallic electrical conductivity discovered so far. Although the lattice thermal conductivity of $Bi_3Ir_3O_{11}$ is close to its minimum limit, the electron thermal conductivity can still be described by the Wiedemann-Franz law with (or close to) a conventional Lorenz constant, $L_0$, regardless of the electronic correlated effects. This work opens a new way to find unprecedented metallic materials with ultralow thermal conductivities, and potentially contributes to the future development of heat-insulation metals.

This work was supported by the National Key R&D Program of China (Grant No. 2021YFA1400300, 2018YFA0305700, 2018YFA0702100), the National Natural Science Foundation of China (Grant No. 11934017, 11921004, 12204516, U1601213), the Beijing Natural Science Foundation (Grant No. Z200007), and the Chinese Academy of Sciences (Grant No. XDB33000000). Z. H. L. acknowledges the support from the China Postdoctoral Innovative Talent program. We acknowledge support from the Max Planck-POSTECH-Hsinchu Center for Complex Phase Materials.

§Corresponding auother.

hzhao@iphy.ac.cn

‡Corresponding auother.

zhangwq@sustech.edu.cn

†Corresponding auother.

ywlong@iphy.ac.cn

*These authors contributed equally to this work

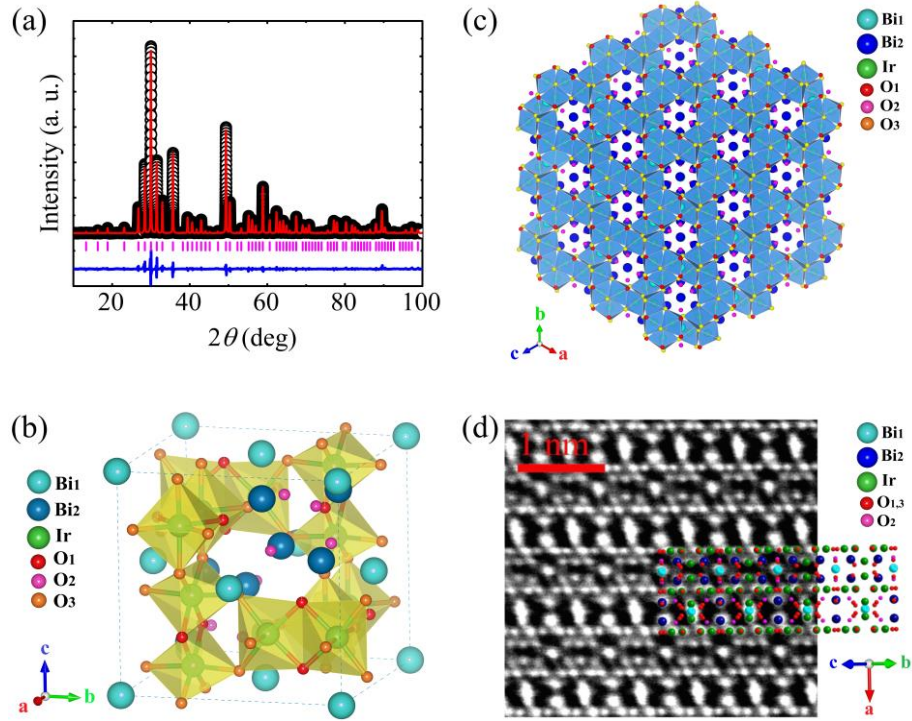

FIG. 1. (a) XRD pattern and structure refinement results of $Bi_3Ir_3O_{11}$. The observed (black circles), calculated (red line), difference (blue line), and allowed Bragg reflections (ticks) with space group $Pn$-3 are shown. (**b**) Schematic structure of $Bi_3Ir_3O_{11}$ composed of corner- and edge-sharing $IrO_6$ octahedra. (**c**) The crystal structure is viewed along [111] direction to demonstrate the tunnel-like frameworks with rattling Bi atoms. (d) HAADF image of $Bi_3Ir_3O_{11}$ along the [110] direction.

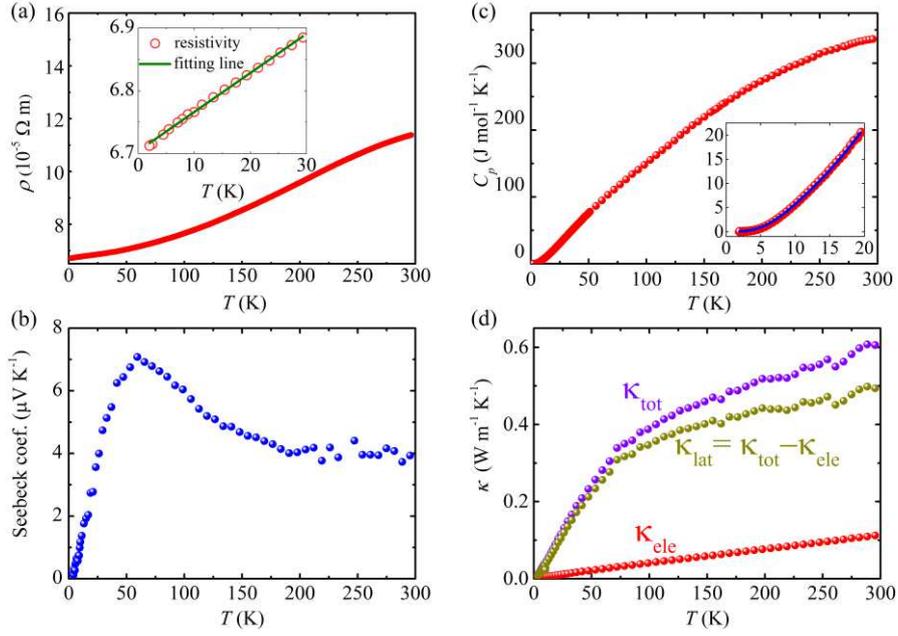

FIG. 2. Temperature dependence of (a) resistivity, (b) Seebeck coefficient, (c) heat capacity, and (d) total thermal conductivity (violet spheres), lattice thermal conductivity (dark yellow spheres), and electron thermal conductivity (red spheres) of $Bi_3Ir_3O_{11}$. The green line in the inset of (a) shows the linear fitting of resistivity below 30 K. The blue curve in the inset of (c) shows the fitting of heat capacity below 20 K.

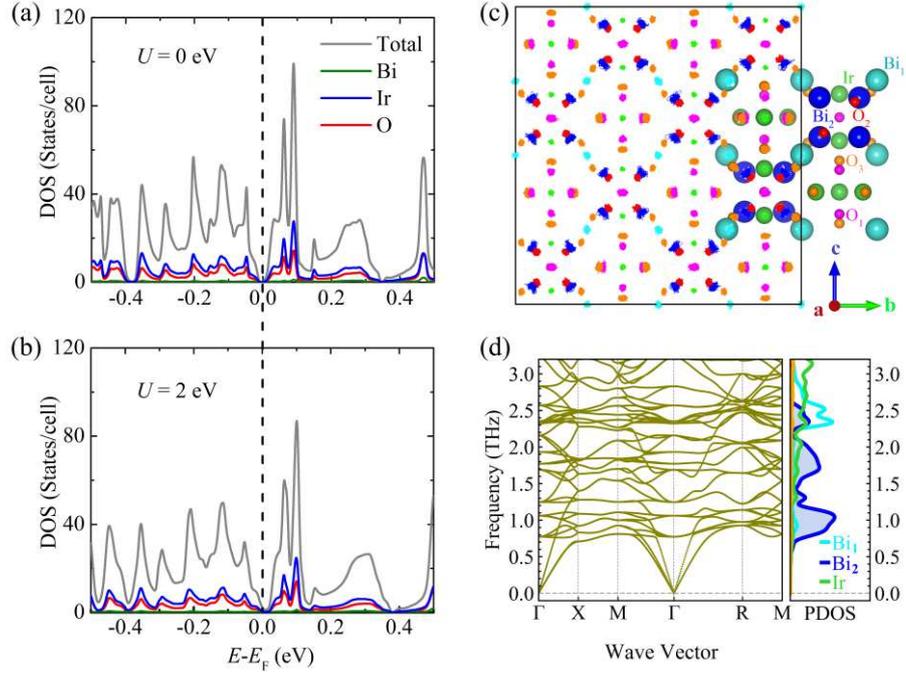

FIG. 3. First-principles calculations with (a) GGA scheme and (b) GGA+$U$ scheme, $U$ = 2 eV. (c) The MD-simulated trajectory of atoms along the (100) direction of $Bi_3Ir_3O_{11}$ at 300 K. $Bi_1$, $Bi_2$, Ir, and $O_1$ are denoted in cyan, blue, olive green, and magenta, respectively. $O_2$ and $O_3$ are both denoted in red. (d) Phonon dispersion relation of $Bi_3Ir_3O_{11}$.

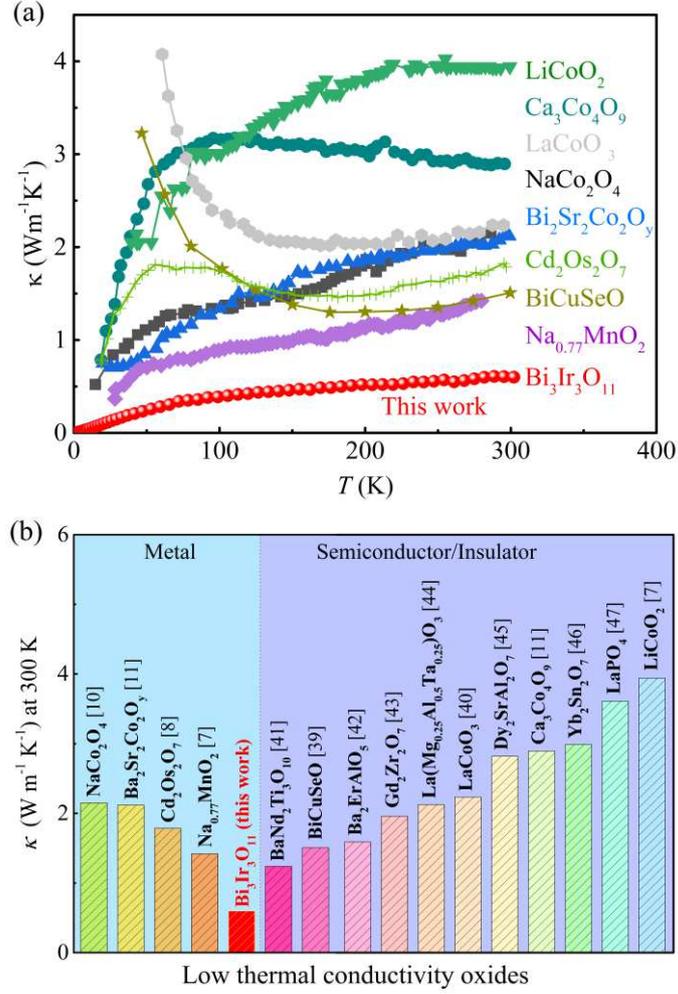

FIG. 4. (a) Thermal conductivity as a function of temperature below 300 K. (b) The values of thermal conductivity determined at 300 K for low-$\kappa_{tot}$ metallic oxides [7,8,10,11] and other more semiconducting/insulating oxides [7,11,39-47].

Supplementary Materials

# **Discovery of a metallic oxide with ultralow thermal conductivity**

Jianhong Dai[1,2,*], Zhehong Liu[1,2,*], Jialin Ji[3], Xuejuan Dong[1], Jihai Yu[4], Xubin Ye[1,2],

Weipeng Wang[1], RiCheng Yu[1,2], Zhiwei Hu[5], Huaizhou Zhao[1,2,§], Xiangang Wan[4],

Wenqing Zhang[6,7,‡], & Youwen Long[1,2,8,†]

[1]Beijing National Laboratory for Condensed Matter Physics, Institute of Physics,

Chinese Academy of Sciences, Beijing 100190, China

[2]School of Physics, University of Chinese Academy of Sciences, Beijing 100049, China

[3]Materials Genome Institute, Shanghai University, Shanghai 200444, China.

[4]Department of Physics, Nanjing University, Nanjing 210093, China

[5]Max Planck Institute for Chemical Physics of Solids, NöthnitzerStraße 40, 01187

Dresden, Germany

[6]Department of Materials Science and Engineering and Shenzhen Institute for Quantum

Science & Technology, Southern University of Science and Technology, Shenzhen,

Guangdong 518055, China.

[7]Shenzhen Municipal Key Lab for Advanced Quantum Materials and Devices and

Guangdong Provincial Key Lab for Computational Science and Materials Design,

Southern University of Science and Technology, Shenzhen, Guangdong 518055, China.

[8]Songshan Lake Materials Laboratory, Dongguan 523808, Guangdong, China

**Methods**

**Sample preparation.** $Bi_3Ir_3O_{11}$ was synthesized by high-pressure and high-temperature techniques using a cubic anvil-type high-pressure apparatus. Stoichiometric $Bi_2O_3$ (> 99.9%) and Ir powders (> 99.9%) were mixed with excess $KClO_4$ as an oxidizing agent. All reactants were placed in an agate mortar and fully ground in an argon-filled glove box. The mixture was then packed into a platinum capsule (3.0 mm diameter and 4.0 mm height), followed by synthesizing at 6.0 GPa and 1373 K for 45 min. After the reaction, the sample was quenched to room temperature, and the pressure was slowly released. Residual KCl was washed out using diluted ionic water.

**Crystal structure analysis.** Powder XRD was performed at room temperature on a Huber diffractometer equipped with Cu-$K_{\alpha1}$ radiation. The XRD data were refined using the Rietveld method with the GSAS program. Scanning transmission electron microscopy (STEM) (HAADF and ABF) and SAED were performed using a JEM-ARM200F microscope with double Cs correctors for the condenser and objective lenses. The available point resolution was better than 0.78 Å at an operating voltage of 200 kV. The acceptance angles were 90–370 mrad for the HAADF imaging and 10-20 mrad for the ABF imaging. All images shown in this text were Fourier-filtered to minimize the effect of the contrast noise.

**Electrical, thermal, magnetic and spectral properties measurements.** Washed powder samples were pressed into dense pellets at 6 GPa and room temperature to measure their physical properties. The resistivity, heat capacity, Seebeck coefficient, and thermal conductivity were measured on the same pellet in the temperature range of 2-300 K using a physical property measurement system (Quantum Design, PPMS-9T). Magnetic susceptibility and magnetization were measured using a commercial superconducting quantum interference device magnetometer (Quantum Design, MPMS-7 T). The ultrasonic measurements of the longitudinal and transversal phonon velocities were performed at room temperature using a Panametrics NDT 5077PR squarewave pulser/receiver head with a Tektronix TDS 2012C digital oscilloscope recording the response. X-ray absorption (XAS) was measured at BL 07A of NSRRC in transmission geometry.

**Electronic structure calculations.** The electronic band structures and density of states (DOS) calculations were performed using the full potential linearized augmented plane-wave method, as implemented in the WIEN2K package [1]. The generalized gradient approximation (GGA) of Perdew-Burke-Ernzerhof (PBE) was employed in our calculation to address the exchange and correlation among the localized $d$ electrons [2]. A $10 \times 10 \times 10$ k-point mesh was used for the Brillouin zone integral. Using the second-order variational procedure, we included spin-orbit coupling (SOC) [3], which has been found to play an important role in the $5d$ system. Self-consistent calculations are considered to converge when the difference in the total energy of the crystal does not exceed 0.1 mRy. We utilized the GGA + $U$ scheme [4] to consider the effect of Coulomb repulsion in the $5d$ orbitals. We varied the parameter $U$ between 0.0 and 5.0 eV.

**Phonon dispersion calculations.** The phonon spectrum was obtained by *ab* initio molecular dynamics (AIMD) based on the Vienna *ab* initio simulation package (VASP). For structure optimization, summation over the Brillouin zone (BZ) was performed with a $4 \times 4 \times 4$ Monkhorst–Pack $k$ mesh for the primitive cell. The plane-wave energy cut-off was set to 520 eV. A convergence criterion of $10^{-5}$ eV of the Hellmann–Feynman force and convergence criterion of the energy convergence criterion of $10^{-8}$ eV were constructed for calculation. Molecular dynamics simulations were carried out in a canonical ensemble (NVT) with a constant volume and temperature based on the Nosé–Hoover thermostat. A supercell with 544 atoms ($2 \times 2 \times 2$ of the unit cell) was used for the calculations. A plane-wave energy cut-off of 300 eV was selected for the calculation. The convergence criteria of the Hellmann–Feynman force and the energy convergence criterion were $10^{-4}$ eV. The simulation temperature was set to 300 K and the simulation time was longer than 20 ps (with a time step of 1 fs) to ensure convergence.

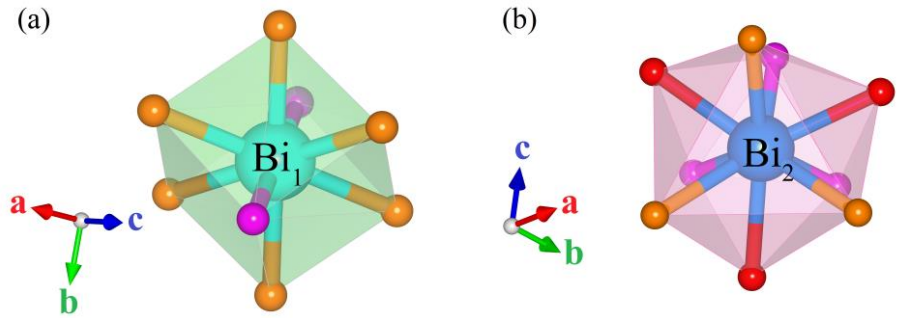

FIG. S1. Schematic diagram of (a) $Bi_1$ atom coordinated by 8 oxygen atoms. (b) $Bi_2$ atom coordinated by 9 oxygen atoms.

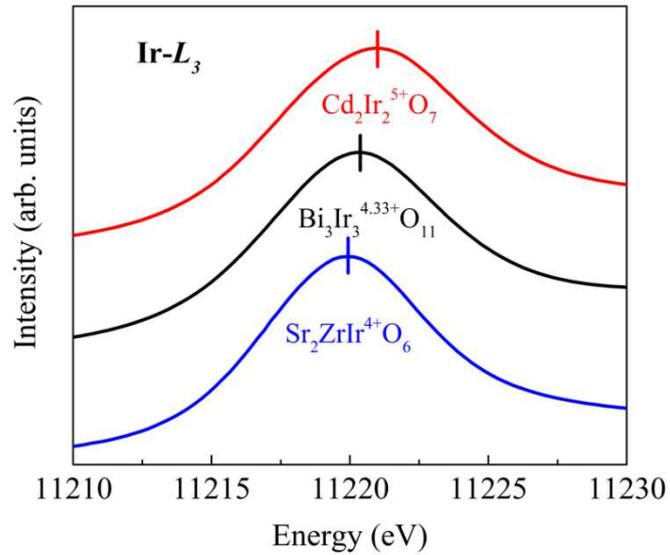

FIG. S2. X-ray absorption spectroscopy of the Ir-$L_3$ edges of $Bi_3Ir_3O_{11}$. $Sr_2ZrIrO_6$ and $Cd_2Ir_2O_7$ [5] were used as $Ir^{4+}$ and $Ir^{5+}$ references with $IrO_6$ octahedral coordination, respectively. According to the energy positions of $Sr_2ZrIrO_6$ (11220.00 eV) and $Cd_2Ir_2O_7$ (11221.00 eV), a mixed valence state of +4.33 is assigned to Ir for $Bi_3Ir_3O_{11}$.

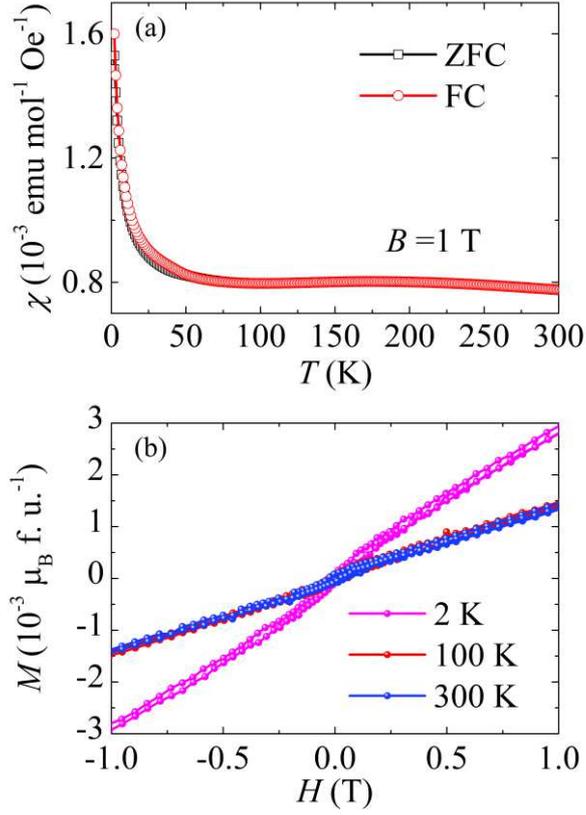

FIG. S3. (a) Temperature dependence of magnetic susceptibility measured at 1 T in zero-field-cooling (ZFC) and field-cooling (FC) mode for Bi$_3$Ir$_3$O$_{11}$. (b) Field dependent magnetization measured at 2 K, 100 K and 300 K.

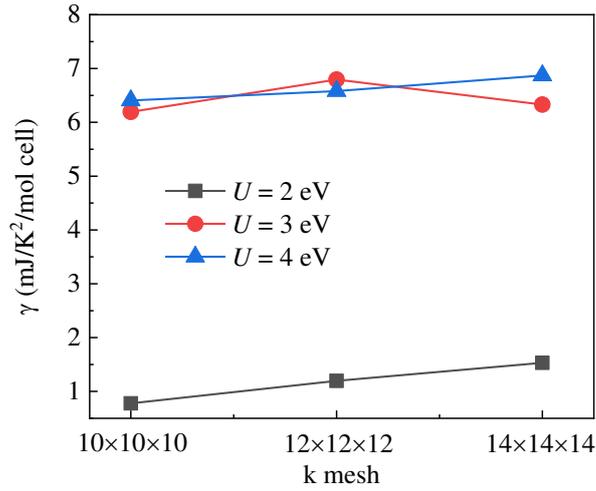

FIG. S4. Calculated Sommerfeld coefficient γ at different k mesh. Different electron-electron correlated energy $U$ ($U$ = 2, 3, 4 eV) are used for Bi$_3$Ir$_3$O$_{11}$.

Table S1: Atomic position of $Bi_3Ir_3O_{11}$ refined based on the Rietveld method using the GSAS program.

| Atom | Site | x | y | z | Occupancy |
|------|------|---|---|---|-----------|
| $Bi_1$ | 4b | 0 | 0 | 0 | 0.982(6) |
| $Bi_2$ | 8e | 0.3827(1) | 0.3827(1) | 0.3827(1) | 0.993(6) |
| Ir | 12g | 0.4072(1) | 0.75 | 0.25 | 0.997(6) |
| $O_1$ | 12f | 0.6050(8) | 0.25 | 0.25 | 1.0 |
| $O_2$ | 8e | 0.1345(6) | 0.1345(6) | 0.1345(6) | 1.0 |
| $O_3$ | 24h | 0.5822(6) | 0.2473(5) | 0.5490(7) | 1.0 |

Table S2: Selected structural parameters of $Bi_3Ir_3O_{11}$ refined based on the Rietveld method using the GSAS program.

| Space group | $Pn$-3 |
|-------------|--------|
| a (Å) | 9.4110(1) |
| $Bi_1$-$O_2$ ×2 (Å) | 2.237(5) |
| $Bi_1$-$O_3$ ×6 (Å) | 2.490(5) |
| $Bi_2$-$O_1$ ×3 (Å) | 2.746(6) |
| $Bi_2$-$O_2$ ×3 (Å) | 2.324(5) |
| $Bi_2$-$O_3$ ×3 (Å) | 2.739(6) |
| Ir-$O_1$ ×2 (Å) | 2.006(5) |
| Ir-$O_3$ ×2 (Å) | 2.042(6) |
| | 1.930(7) |
| Ir-$O_1$-Ir (°) | 95.1(3) |
| Ir-$O_3$-Ir (°) | 126.6(3) |
| $R_p$ | 5.0% |
| $R_{wp}$ | 7.3% |

§Corresponding auother.

hzhao@iphy.ac.cn

‡Corresponding auother.

zhangwq@sustech.edu.cn

†Corresponding auother.

ywlong@iphy.ac.cn

*These authors contributed equally to this work.